**Bistability of the $B_iO_i$ complex and its implications on evaluating the "acceptor removal" process in p-type silicon**

C. Besleaga[1,*], A. Kuncser[1,*], A. Nitescu[1], G. Kramberger[2], M. Moll[3] and I. Pintilie[1,**]

[1] National Institute of Materials Physics, Atomistilor 405A, Magurele 077125, Romania

[2] Jozef Stefan Institute, University of Ljubljana, 100 Ljubljana, Slovenia

[3] CERN, 1211 Geneva, Switzerland

**Abstract**

The dependencies of the $B_iO_i$ defect concentration on doping, irradiation fluence and particle type in p-type silicon diodes have been investigated. We evidenced that large data scattering occurs for fluences above $10^{12}$ 1 MeV neutrons/cm$^2$, becoming significant larger for higher fluences. We show that the $B_iO_i$ defect is metastable, with two configurations A and B, of which only A is detected by Deep Level Transient Spectroscopy and Thermally Stimulated Currents techniques. The defect' electrical activity is influenced by the inherent variations in ambient and procedural experimental conditions, resulting not only in a large scattering of the results coming from the same type of measurement but making any correlation between different types of experiments difficult. It is evidenced that the variations in $[B_iO_i^A]$ are triggered by subjecting the samples to an excess of carriers, by either heating or an inherent short exposure to ambient light when manipulating the samples prior to experiments. It causes ~7h variations in both, the $[B_iO_i^A]$ and in the effective space charge. The analyses of structural damage in a diode irradiated with $10^{19}$ 1 MeV neutrons/cm$^2$ revealed that the Si structure remains crystalline and vacancies and interstitials organize in parallel tracks normal to the Si-SiO$_2$ interface.

[*] Equal contribution as first authors

[**] corresponding author: ioana@infim.ro

Keywords: metastable boron-oxygen defect, bistability, acceptor removal rate, DLTS, TSC, HRTEM

# 1. Introduction

The detectors presently installed in the Large Hadron Collider (LHC) at CERN are essentially based on n-type silicon and are designed to withstand an integrated luminosity of 300 fb$^{-1}$, corresponding to cumulated radiation levels of 2×10$^{15}$ n$_{eq}$/cm$^2$ (n$_{eq}$ stands for equivalent 1 MeV neutrons) and ionizing doses of 300 kGy. For the upgrade of the LHC, the High-Luminosity LHC (HL-LHC) expected to operate at the end of 2027, a more than 10 times increase in the integrated luminosity (~4000 fb$^{-1}$) and in the corresponding radiation levels (~3x10$^{16}$ n$_{eq}$/cm$^{2)}$ is anticipated [1]. To face such harsh radiation environments while assuring proper performance, new generations of silicon sensors with improved timing performance and low cost large area detector technologies have to be developed. The selected choice for the ATLAS and CMS Strip Tracker upgrades for HL-LHC is to use devices based on p-type silicon, a solution brought forward and tested by the RD50 community [2]. Accordingly, various type of sensors based on boron-doped silicon, started to be developed and studied, from simple n$^+$- p strip and pixel silicon to sensors with intrinsic gain for precision timing (e.g. Low Gain Avalanche Detectors - LGADs) [3-5] and monolithic sensors able to amplify the signal within the pixel cell (e.g. High Voltage CMOS sensors) [6,7]. Radiation effects in these devices were surveyed mainly at macroscopic level, by monitoring their electrical performance after irradiation with different particles of various fluence levels and the changes during annealing at moderate temperatures (60, 80 °C) [8-11]. The latter is important for planning running scenarios as well as for quantitatively predicting the effects of the annealing on operation if unplanned events occur. Common for all these new sensors is the loss of initial boron doping concentration caused by irradiation, a process known as "acceptor removal" [12,13]. The decrease induced by irradiation on the effective doping concentration (N$_{eff}$) is evaluated from the depletion voltage V$_{dep}$ resulted from Capacitance-Voltage (C-V) characteristics measured usually at temperatures between -20°C and +20°C and for an a.c. frequency of 10 kHz. If for n-type sensors, the change in N$_{eff}$ due to irradiation and annealing could be quantitatively explained based on the results of defect investigations after many years of research [14-16], for boron-doped p-type devices the defect investigation research is less advanced in succeeding to quantitatively explain the changes observed in their C-V characteristics caused by irradiation. One of the main reasons for the "acceptor removal" process in boron-doped irradiated silicon is the generation of the B$_i$O$_i$ complex with a donor level located at ~ 0.25 eV from the conduction band of silicon. Such a defect provides positive space charge in the depletion region of silicon diodes.

Thus, for each $B_iO_i$ defect created, one shallow acceptor ($B_s$) is removed and one charged donor is formed. The $B_iO_i$ donor energy level can be easily detected by Deep Level Transient Spectroscopy (DLTS) and Thermally Stimulated Currents (TSC) technique after charge injection with forward biasing the diodes. The introduction rate of a defect D is defined as the ratio between the defect concentration and the received irradiation fluence ($\Phi$), g=[D]/$\Phi$ [17]. Regarding the introduction rate of the $B_iO_i$ defect, $g_{BiOi}$, although the reported data follows the expected increasing trend with the initial B-doping, they do not match the values of the "acceptor removal" rate – $g_B$ as measured from C-V characteristics, $g_{BiOi}$ being smaller than $g_B$ with a factor varying randomly between 2 and 4 [13, 18]. In addition, the collected $g_B$ data in Ref. 13 shows a large scattering, which requires further investigations in order to obtain a better understanding. The consequence of the latter has a major impact on operation of detectors exploiting gain in highly doped p+ layer and explains the very large differences in charge collection and timing performance of the same devices irradiated to the same fluence - see [19] and the references therein. According to the Watkins replacement mechanism [20], an important competing reaction to the formation of the $B_iO_i$ defect, involves the Carbon impurity in B-doped silicon. Thus, the interstitials generated by irradiation can react with both type of substitutional atoms, boron or Carbon, moving the corresponding B and C atoms on interstitial sites according to the reactions I+$B_s$→$B_i$ and I+$C_s$→$C_i$, respectively. The interstitials $B_i$ and $C_i$ migrating through the lattice, can further react with Oxygen interstitials ($O_i$), which are always present in the silicon crystal, leading to the formation of $B_iO_i$ and $C_iO_i$ defect complexes ($B_i$+$O_i$→$B_iO_i$, $C_i$+$O_i$→$C_iO_i$) that are stable at ambient temperatures. Dependencies on B and C content in the samples, irradiation fluence and type of the incident particle are thus, expected reasons for the variations of $g_B$ and $g_{BiOi}$. However, they alone, do not explain the large spread of the reported data. By this "scattering" our research efforts were triggered, trying to understand the unusually large spread of the results on boron doped silicon samples. We succeed in showing that the scattering is not necessarily due to measurement errors or material variations but can easily be the consequence of the inherent normal manipulation of the samples. In the present work we will show that what appears to be a random large scattering of the data can be justified by a valid physical explanation - the bistable nature of the $B_iO_i$ defect.

As the high-energy physics experiments are more and more demanding, the damage induced by irradiation in the crystalline structure of the Si sensors is also an ongoing subject of studies. There are numerous examples in the literature, investigating different facets of the structural damage problem [21-27]. However, the considered fluences did not exceed $10^{15}$ $n_{eq}$/cm$^2$ We will present the results of structural investigations performed on LGAD samples irradiated with 1 MeV neutrons /cm$^2$ at an extremely high fluence, of $10^{19}$ cm$^{-2}$.

## 2. Materials and Methods

The p-type samples investigated in this work are grouped below in three categories, corresponding to the type of experiments performed after different levels of irradiation:

*A) Irradiation fluences between $10^{10}$ and $10^{13}$ $n_{eq}$/cm$^2$ – DLTS investigations*

We have investigated a serie of Si n$^{++}$-p-p$^{++}$ pad diodes, 2.632 × 2.632 mm², processed on 45 µm thick epitaxial grown Si layer having as substrate 10 Ωcm Czochralski (Cz) -grown silicon. The growth of all of the epilayers was performed by ITME Warsaw, Poland. The only varying parameter imposed for the present study was the amount of incorporated boron doping during the epitaxial growth. The epitaxial grown (EPI) diodes have been produced by CiS Forschungsinstitut für Mikrosensorik GmbH, Erfurt, Germany, using the same implant and processing procedures. This way, diodes of different resitivities were obtained, of 50, 250 and 1000 Ωcm, all having the same amount of C and O impurities ([C] ~ 1.3×10$^{16}$ cm$^{-3}$ and [O] ~ 3×10$^{17}$ cm$^{-3}$). All diodes had guard rings that in all of our experiments were grounded. Four of each type of diodes were investigated as follows: 2 after irradiation with 1 MeV neutrons (2 fluences) and 2 after irradiation with 23 GeV protons (irradiated to the same equivalent fluences as the irradiation with 1 MeV neutrons). The EPI diodes are labelled in the following according to their bulk resistivity as EPI-50, EPI-250 and EPI-1000 while the type of irradiation will be denoted with „n" or „p". The 23 GeV proton irradiation fluence will be given as the equivalent to 1MeV neutron irradiation fluence and will be denoted by $\Phi_{eq}$ while for irradiation with neutrons the „eq" subscript will be ommited. For determining $\Phi_{eq}$ a hardness factor of 0.62 was considered [28]. These pad diodes were investigated by means of Deep Level Transient Spectroscopy (DLTS), after injecting only holes or both electrons and holes with filling pulses ranging from 1 ns to 10 s.

By using same vendors and processing of all diodes while performing *all the experiments in the same place*, with the same set-up/procedures the study performed on EPI diodes becomes relevant for studying the dependencies of $g_{BiOi}$ on boron doping, on fluence and type of irradiation, for samples with the same Carbon content. It is worth noting that the DLTS technique is applicable when the defect concentrations do not exceed 10% of the doping concentration in the samples and the possible variations induced by defects in C-V characteristics at ambient temperatures are insignificant.

*B) Irradiation fluences of $10^{14}$ and $10^{15}$ $n_{eq}/cm^2$ – TSC investigations*

A pair of samples, consisting of B-doped 45 µm thick $n^{++}$-p-$p^{++}$ pad diodes, 3×3 mm$^2$, produced on the same high resistivity standard float-zone (STFZ) wafer by Centro Nacional de Microelectrónica, Barcelona, Spain (CNM), have been investigated in this study. These samples will be labelled further on as STFZ. Both diodes were measured with the guard rings grounded. The PAD samples were irradiated with 1 MeV neutrons at the Triga Reactor in Ljubljana, with a fluences of $10^{14}$ and $10^{15}$ n/cm$^2$. After such irradiation levels, the bulk of the samples is compensated and the DLTS technique, based on measuring high-frecquency signals, is not applicable anymore. This is why also the evaluations from the 10 kHz C-V characteristics are less reliable. For these samples we have monitored the variations in the depletion voltage via d.c. Current-Voltage (I-V) characteristics. We have made defect investigations by employing the TSC technique using the procedures described in more detail in [29, 30]. With TSC method a d.c. current generated by the charge emitted from the defects is measured and so, the technique is applicable on highly irradiated samples where methods based on high-frequency signals cannot be used anymore. The TSC technique was successfully applied previously on highly irradiated n-type silicon sensors, revealing the radiation induced defects quantitatively explaining the changes in $N_{eff}$ caused by irradiation and annealing of Phosphorous – doped silicon sensors [15, 16].

As it will be described later on, the defect investigations performed on samples exposed to such levels of irradiation will point out some procedural reasons leading to the detection of a variation in both, $N_{eff}$ as determined from I-V d.c. measurements and introduction rates of the $B_iO_i$ defect as evaluated from TSC spectra.

*C) Irradiation fluence of $10^{19}$ 1 MeV neutrons/cm$^2$ – microstructural investigations*

A highly irradiated LGAD sample was subjected to micro-structural investigations. Provided the extreme fluence considered for the investigated sample, transmission electron microscopy (TEM) is used for structural investigations and defects (e.g. vacancies, interstitials, extended defects) that are expected to be revealed in the Si lattice. As TEM observations involve electron irradiation of the investigated sample, precautions need to be taken in order to avoid electron-induced artifacts. Threshold values for Frenkel defects generation in Si are considered to be in the range 175 keV to 260 keV, which includes the widely used 200 keV energy of the TEMs [31, 32]. As the authors in Ref. 33 report that even severe events such as amorphization could take place in thin Si under microscope observation, it is clear that the problem of electron induced artifacts is a complex one. R.F. Egerton *et al* indicates in some guidelines in avoiding beam induced artifacts which have been considered when performing the experiments described in this paper [34]. We have used a JEOL 2100 Transmission Electron Microscope equipped with LaB6 electron gun and high resolution pole piece. For in-situ investigations the JEOL heating control unit has been used together with a JEOL heating holder in order to precisely control the applied temperature. The microscope has been operated at 200 kV. The samples have been prepared using the conventional cross-section method (X-TEM) based on mechanical thinning and Ar ion polishing. In order to avoid exposing the material to heat during the sample preparation, the glue of the X-TEM specimen has been dried for 2 days at room temperature before the ion polishing. The sample has been investigated by HRTEM (High Resolution Transmission Electron Microscopy) and by SAED (Selected Area Electron Diffraction).

## 3. Results and Disscusions

In the following we will present the results of our investigations on the samples described above. The defect investigations were always accompanied by C-V/I-V measurements at 293K, aiming to understand the real impact the detected defects have on the leakage current (LC) and $N_{eff}$. When the concentration, charge states (donor or acceptor), activation energy and both capture cross sections of a defect are known then its impact on LC and $N_{eff}$ of the diodes can be calculated as function on temperature based on Shockley–Read–Hall statistics [11, 35, 36]. For p-type diodes and low densities of free carriers in the space charge region of the diodes, the variations on LC and $N_{eff}$ induced by a defect can be calculated according to the following equations:

$$\Delta LC(T) = q \times A \times d \times N_t \frac{e_p(T) \times e_n(T)}{e_n(T) + e_p(T)} \quad (1)$$

$$\Delta N_{eff}^{acceptor}(T) = N_t^{acceptor} \frac{e_p(T)}{e_n(T) + e_p(T)} \quad (2)$$

$$\Delta N_{eff}^{donor}(T) = -N_t^{donor} \frac{e_n(T)}{e_n(T) + e_p(T)} \qquad (3)$$

$$\text{with } e_{n,p} = v_{n,p}^{th} \times \sigma_{n,p}(T) \times N_{C,V} \times \exp\left(\mp \frac{E_{C,V} - E_a}{k_B \times T}\right)$$

Where q is the elementary charge, A is the area of the diode, d is the diode thickness, $v^{th}_{n,p}$ are the thermal velocities of electrons (n) or holes (p), $N_t$ is the defect concentration, $N_{C,V}$ are the effective density of states in the conduction/valence band, $E_C$ and $E_V$ are the conduction and valence band edge energies, respectively.

*A) Irradiation fluences between $10^{10}$ and $10^{13}$ $n_{eq}/cm^2$ – DLTS investigations*

Typical DLTS spectra obtained on the series of EPI diodes of different resistivity are shown in Figure 1 and Figure 2 for samples irradiated with 1 MeV neutrons and 23 GeV protons, respectively. The samples were cooled under zero bias and all the DLTS measurements were performed during heating from 30 K to 290 K. The measuring conditions, reverse bias ($U_R$), pulse bias ($U_P$), pulse duration ($t_P$) and time window ($T_w$) are given in the figures. The recorded DLTS peaks are labelled according to the defect identities and charge states of the detected energy levels for known defect levels or to the peak temperatures for yet un-identified traps. The known energy levels giving rise to peaks in DLTS measurements are coming from $I_2O^{+/0}$, $V_2^{+/0}$, $V_3^{+/0}$ and $C_iO_i^{+/0}$ in spectra evidencing the hole traps and from $B_iO_i^{0/+}$ in spectra corresponding to injection of electrons and holes [12, 18, 37-43]. The most prominent un-identified energy levels in the spectra for which we could accurately determine the trapping parameters are denoted as H156K and H223K and they are trapping holes.

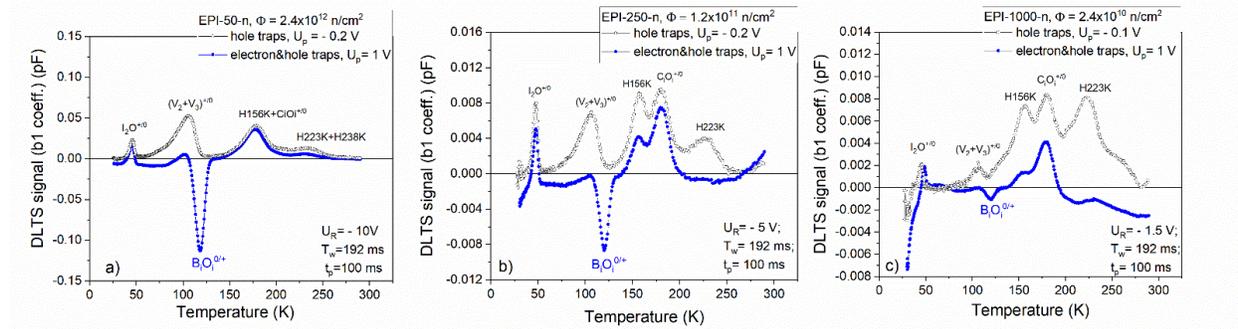

**Figure 1.** DLTS spectra after injecting only majority carriers (hole traps) or both type of cariers (electron and hole traps) measured on diodes of different resistivity after irradiation with 1 MeV neutrons: a) EPI-50-n; b) EPI-250-n; c) EPI-1000-n.

For all well separated signals we also performed direct measurements of the capture cross sections in the temperature range of the corresponding DLTS peaks. The trapping parameters of H156K and H223K energy levels were determined from measurements performed on EPI-250 and EPI-1000 diodes. The values of the trapping parameters obtained from direct measurements of the capture cross sections at the corresponding DLTS peak temperatures are: $E_a^{H156K} = E_v + 0.291$ eV, $\sigma_p^{H156K} = 4.8 \times 10^{-16}$ cm$^2$ and $E_a^{H223K} = E_v + 0.363$ eV, $\sigma_p^{H223K} = 1.7 \times 10^{-17}$ cm$^2$.

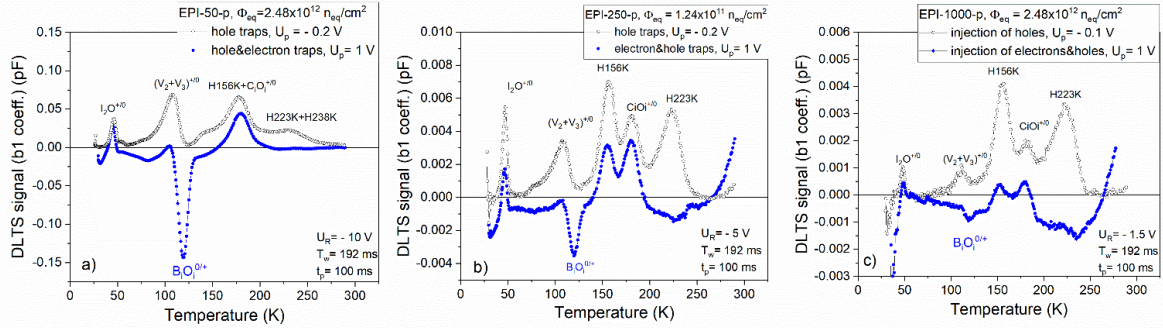

**Figure 2.** DLTS spectra after injecting into defects only majority carriers (hole traps) or both type of carriers (electron and hole traps) measured on diodes of different resistivity after irradiation with 23 GeV protons: a) EPI-50-p; b) EPI-250-p; c) EPI-1000-p.

For the $B_iO_i^{0/+}$ energy level, which is the primary focus of the present study, we measured both capture cross sections, for electrons ($\sigma_n$) and for holes ($\sigma_p$) at the corresponding DLTS peak temperature of 120 K. Examples of such measurements are given in Figure 3. Accurate values of $\sigma_n$ and $\sigma_p$ were obtained from the measurements on the diodes of 250 Ωcm resistivity, where the $B_iO_i$ defect provided a good DLTS signal and the necessary small enough pulses for determining $\sigma_n$ could be generated by our DLTS system.

For $\sigma_p$ we have used the double pulse method proposed by Lang [44] consisting in a sequence of two consecutive injection pulses, first with forward bias, assuring the filling of $B_iO_i$ defect with electrons, followed by the second one, injecting only holes.

The measured data together with the determined values for $\sigma_n$ and $\sigma_p$ are given in Figure 3. The activation energy was then determined to vary between 0.245 eV and 0.251 eV from the conduction band for electric fields between 16.2 kV/cm and 1.61 kV/cm, a result in agreement with the expected behavior for a donor in the upper part of the band gap.

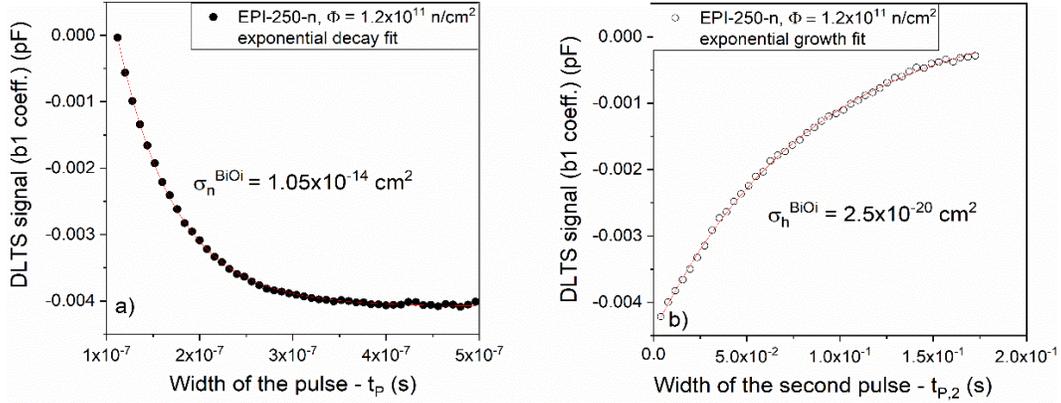

**Figure 3.** Direct measurement of the capture cross sections ($\sigma_{n,p}$) of the $B_iO_i^{0/+}$ defect energy level at 120K: a) $\sigma_n$, after injection pulses of $U_P = 1$ V and different pulse widths; b) $\sigma_p$, after double injection pulses, first with $U_{P,1} = 1$ V, $t_P = 100$ ms followed by the second one with $U_{P,2} = -0.2$ V.

With both capture cross sections being experimentally determined for the $B_iO_i$ defect and with the donor level seen in DLTS spectra as a minimum at ~ 120 K, it was calculated using Eqs. 1 and 3, that the defect majorily impacts $N_{eff}$ at ambient temperatures, the $B_iO_i$ contributing in its full concentration (100%) with positive charge, while it is negligible on LC (e.g. a $B_iO_i$ concentration of $10^{13}$ cm$^{-3}$ will generate a 2 pA current at 293 K).

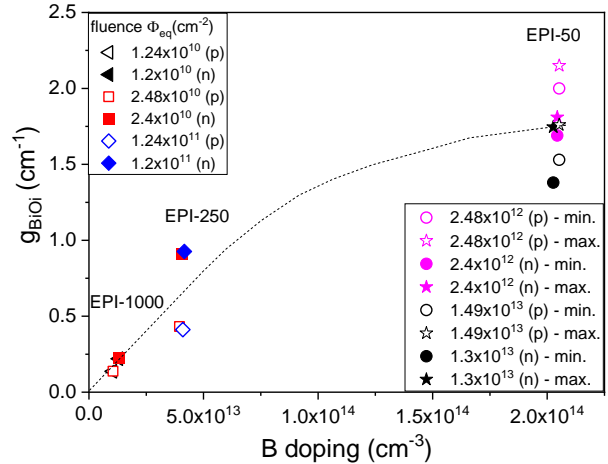

**Figure 4.** All EPI diodes, dependence of $g_{BiOi}$ on B doping, as deduced by C-V measurements.

The dependence of the $B_iO_i$ introduction rate on the boron doping concentration, as deduced by C-V measurements, is given in Figure 4. A non-linear dependence on boron content was reported earlier by Vines et al [45]. One can notice in Figure 4 that only for low boron content, when samples were irradiated with fluences below $2\times10^{11}$ n/cm$^2$, the introduction rates of B$_i$O$_i$ are nicely grouped according to the type of irradiation (neutrons or protons). Repetitive DLTS measurements on samples EPI-1000 and Epi-250, irradiated below $1.2\times10^{11}$ cm$^{-2}$, show no variation in the concentration of B$_i$O$_i$ defect. This is not the case of EPI-50 samples, irradiated with fluences in the $10^{12} - 10^{13}$ cm$^{-2}$ range, where the g$_{BiOi}$ data becomes irreproducible when the samples with the same history behind in terms of irradiation and annealing stage are re-measured. In these low resistivity diodes we detect changes of up to 23% for the detected concentration of B$_i$O$_i$ defect. In the same time, much smaller variations (below 5%) are recorded for the other defects giving rise to peaks in DLTS spectra. We have noticed that the variations in B$_i$O$_i$ can be triggered by both, the thermal treatment performed at 60 °C or simply by a short exposure of the samples to ambient light, inherent in manipulating the samples prior to the measurements. In this respect, we have performed experiments ensuring that all procedural aspects are well controlled. Thus, prior to DLTS experiments, we have left the EPI-50 samples for 15 minutes exposed to the ambient laboratory neon light and consecutive DLTS measurements started imediatly after.

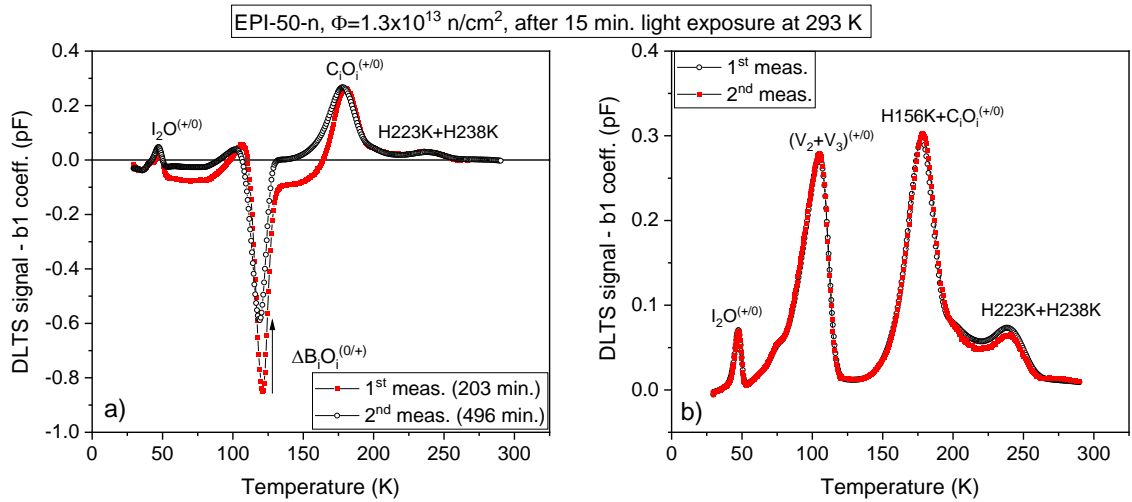

**Figure 5**. Two consecutive DLTS spectra recorded on irradiated EPI-50 diode after exposing the sample to the ambiental laboratory light at 293K for 15 min.: a) electrons&holes traps (U$_P$ = 1 V), the time passed from the illumination at room temperature and the recording of B$_i$O$_i$ peak is given in the legend; b) majority carriers (U$_P$ = 0 V). T$_W$ = 192 ms, t$_p$ = 100 ms and U$_R$ = -10 V for all measurements.

In Figure 5 are given two consecutive DLTS spectra for an EPI-50-n sample recorded after the diode was exposed to 15 min. ambient neon light. Variations in the $B_iO_i$ concentration are observed only between the first and the second DLTS measurement. According to Figure 5 the excess of carriers (during thermal treatment or exposure to light) enhances the DLTS signal from $B_iO_i$ defect. The earliest time after which we can record the DLTS peak at 120 K is 203 min. It takes another ~ 3 hours to record the $B_iO_i$ peak in the second DLTS measurement. In the second DLTS measurement the $B_iO_i$ peak decrease and stabilize to a smaller value than in the first measurement. Repeating the experiment with the light exposure similar variations of $\Delta B_iO_i = 4.2\times10^{12}$ cm$^{-3}$ occur. Such a behaviour suggests a bistable nature of the $B_iO_i$ defect. For EPI-50 samples, we indicate in Figure 4 the minimum and maximum values of $g_{BiOi}$ as determined from repeated DLTS measurements after the last heat treatment performed at 60 °C for 320 minutes (accumulated total treatment of 640 min).

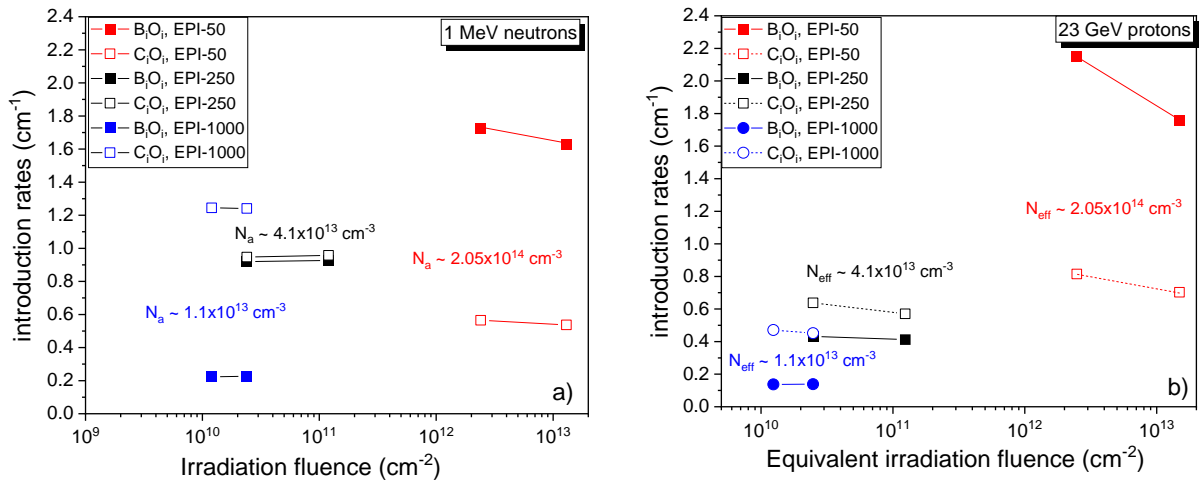

**Figure 6**. All EPI diodes, dependence of $B_iO_i$ and $C_iO_i$ introduction rates on irradiation fluence: a) 1 MeV neutron; b) 23 GeV proton irradiation.

The data in Figures 6a and 6b show also that, after both type of irradiation, $g_{BiOi}$ and $g_{CiOi}$, although different according to the boron doping of the diodes and type of irradiating particle, remain constants only for fluences up to $2\times10^{11}$ n/cm$^3$. Increasing the fluence above this value, a decrease in the introduction rate of both defects is measured, more abrupt for $B_iO_i$.

In Figures 6a and 6b are plotted separately the dependencies on irradiation fluence of the stabilized values determined for $g_{BiOi}$ (filled symbols) from DLTS measurements, after 1 MeV neutrons and 23 GeV protons, respectively. It can be observed that for fluences above $10^{12}$ n$_{eq}$/cm$^2$ the introduction rate of $B_iO_i$ became larger for charged hadrons than for neutrons. The figures contain also the introduction rates of $C_iO_i$ defect (open symbols), the main competitor of the $B_iO_i$ complex. The competition in the formation of these two defects is nicely observed after both type of irradiations, the $B_iO_i$ and $C_iO_i$ trapping centers being generated in similar quantities in EPI-250 diodes. For EPI-1000 the $C_iO_i$ defect becomes dominant while for EPI-50, the $B_iO_i$ center. Both defects are stable at the 60 °C annealing temperature and the only observed stable changes are related to the expected decrease of the $I_2O$ and $V_3$ peak amplitudes (not shown here) [37].

As mentioned above, the bistability of $B_iO_i$ complex was noticed only in EPI-50 diodes, where the $B_iO_i$ is generated in largest amounts among the EPI diodes, reaching values between $5\times10^{12}$ cm$^{-3}$ and $2.6\times10^{13}$ cm$^{-3}$ for irradiation fluences in the $2.4\times10^{12}$ n/cm$^2$ ÷ $1.3\times10^{13}$ n/cm$^2$ range. The variations in $N_{eff}$ caused by the unstable behavior of $B_iO_i$ are below 5% in these low resistivity samples. A better situation in which the noticed bistability of $B_iO_i$ defect can be undoubtfully checked via both, defect characterization and $N_{eff}$ determination, is to investigate high resistivity samples with similar $B_iO_i$ defect concentrations by employing TSC experiments and follow the variations in the depletion voltage of the diodes. Such investigations are described in the following section.

*B) Irradiation fluence of $10^{14}$ and $10^{15}$ $n_{eq}$/cm$^2$ – TSC investigations*

For this study, we have investigated pairs of irradiated PAD and LGAD diodes irradiated with $10^{14}$ and $10^{15}$ n/cm$^2$ 1 MeV neutrons and annealed for more than 90000 minutes at 80 °C. An annealing temperature of 80 °C (or lower) is relevant for the High Energy Physics Community because it preserves the type of the defects induced by irradiation in LHC experiments at CERN and it only accelerates the defect reactions happening at ambient temperatures. The defects detected in samples annealed for so long time are all thermally stable. For this type of samples, possible variations in the determined defect introduction rates caused by procedural factors only, can be more accurately followed. In this respect, it is important to emphasize how the samples are stored and actually measured. The common procedure used for irradiated samples is to firstly ensure that, before and between any kind of measurements and annealing steps, they are stored in freezers at -20 °C, in order to preserve, as accurately possible, the state of the samples. Before starting a measurement, the samples are inherently exposed for a while (not always the same) to the ambient laboratory light. Then, there is an inherent delay between different types of investigations. If for example, the experiments start with measuring C-V/I-V characteristics, which last for about 30 minutes, and then with DLTS/TSC, the time elapsed until the first DLTS or TSC spectrum is measured is usually longer than 2 hours. If the order of measurements reverses, there will be a much longer delay between the moment of exposure to light, the recording of the DLTS/TSC data and of the C-V/I-V measurements. Such variations in the procedures used in investigating various inorganic materials have usually no influence on the measurement results. However, we will show further that for boron-doped Silicon, irradiated with high fluences, such inherent variations in the ambient and procedural conditions result not only in a large scattering of the results coming from the same type of measurement but also makes it difficult to correlate the results of different types of measurements.

Further on we will give examples of two situations, commonly encountered in studies performed on irradiated sensors, the exposure of the samples to heat treatments and to ambient light. The exposure to light and ambiental temperature (that might differ for each laboratory) is inherent when taking the samples out of the freezer or of the oven, as well as just prior to C-V/I-V measurements when in the probe stations a bulb light is switched on for making the electrical contacts. The times the samples are exposed to the ambient conditions are usually not registered nor the time elapsed between a well monitored heat treatment and the start of a specific experiment.

Figure 7a shows the TSC spectra measured on PAD samples just after the heat treatment at 80 °C or exposure to light, along with the repeated ones recorded after the samples were kept 24 h in dark, in the cryostat. Reverse biases of 100 V and 300 V were needed to be applied during the TSC measurements on the PAD samples irradiated with $10^{14}$ and $10^{15}$ n/cm$^2$ in order to fully deplete the structure over the entire temperature range, a needed condition for quantitative analyses [29]. The filling of traps was performed by forward biasing the diodes at 10 K for 30 seconds, the injected current being of 0.3 mA. The first TSC measurement performed on each of the diodes started just after the thermal treatment or exposure to ambient light and the peak corresponding to $B_iO_i$ is recorded in the TSC spectrum 109 minutes after the light exposure or heat treatment of the samples. The second TSC measurement started 24 h later. Between the two measurements the samples were kept in dark in the cryo, at 293K.

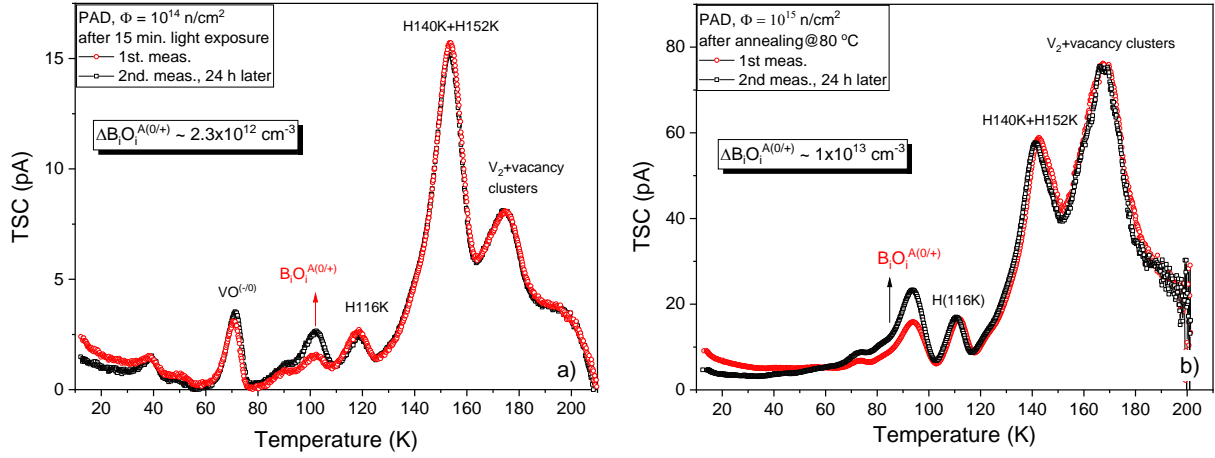

**Figure 7.** TSC spectra on PAD diodes annealed for 91650 min@80 °C. Examples of variations in electrical characteristics and defect concentration caused by heating the diodes to 80 °C (only 30 s exposed to the ambient light) or by exposure to ambient light (laboratory neon for 15 minutes): a) irradiation fluence of $10^{14}$ n/cm$^2$, $V_R = 100$ V; b) irradiation fluence of $10^{15}$ n/cm$^2$, $V_R = 300$ V.

Significant differences between the two measurements are observed only for the $B_iO_i^{0/+}$ energy level. The corresponding TSC peak has a strong increase after keeping the sample a day in the dark, indicating that the defect $B_iO_i$ defect may have at least two different configurations. In the following we will denote the configuration that is manifesting through the electron emission at 100 K with "A" and the other, detected only as variations in the A configuration, with "B". The concentration of $B_iO_i^A$ was obtained by deconvolution of associated TSC peak accounting the formalism described in Ref. 30 and for the Poole Frenkel effect which, due the nonhomogeneous electric field in the diodes, leads to an asymmetry of the corresponding TSC peak [16]. If the donor activity specific to the $B_iO_i^A$ configuration is not preserved when the defect switches from A to B configuration then, the change in the $B_iO_i^A$ concentration should lead to changes in $N_{eff}$ and consequently in the depletion voltage $V_{dep}$. For an abrupt junction and a homogeneous space charge, $V_{dep}$ is given by:

$$V_{dep} = \frac{q|N_{eff}|d^2}{2\varepsilon\varepsilon_0} \quad (4)$$

The I-V characteristics measured at 293K are shown in Figure 8. The $\Delta V_{dep}$ shift is a strong evidence for the variation of $N_{eff}$ in the bulk of the PAD samples caused by light or heat exposure. For both samples the I-V curves are shifting towards smaller reverse bias values, indicating an increase in the concentration of donors during the time spent by the samples in dark. In addition, the total variation in $B_iO_i^A$ concentration as detected in TSC measurements match very well the variation of $N_{eff}$ determined from the I-V measurements, for both of the samples. This means that the configuration $B_iO_i^A$ of the defect is deactivated by an excess of carriers generated either by ambient light or by heating at 80 °C, and the $\Delta[B_iO_i^A]$ amount is redistributed in time from the light/temperature activated B configuration to the stable A state. These variations are measured again if exposures to light or to heat treatments are repeated, clearly indicating the bistable nature of the defect. The steady state, giving the maximum concentration of the $B_iO_i$ in the A configuration, is fully recovered by keeping the sample some time in the dark.

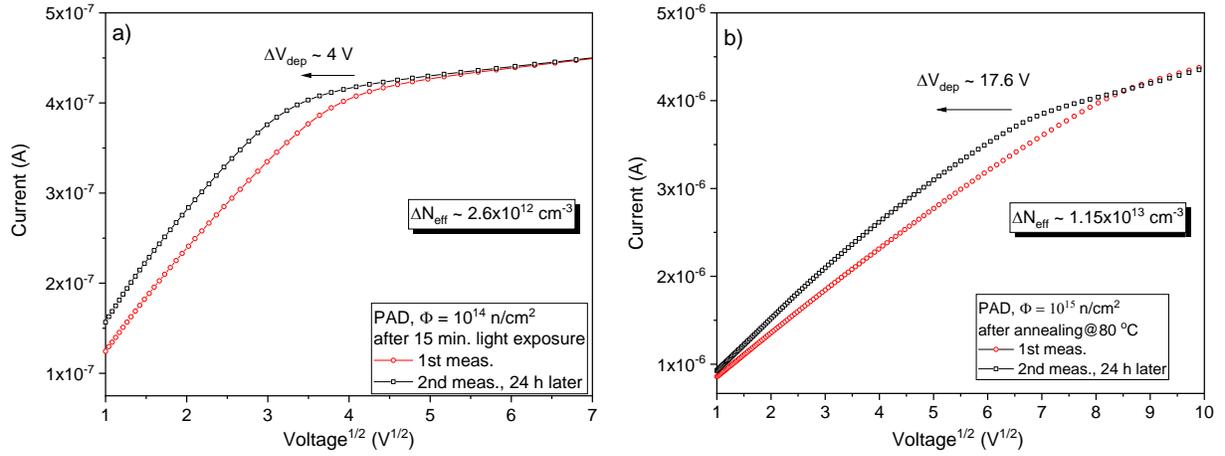

**Figure 8.** Sets of I-V characteristics performed at 293K, 1$^{st}$ after exposing the samples to an excess of carriers (via thermal treatment at 80 °C or exposure for 15 min. to ambient light) on neutron irradiated PAD diodes and fluences of: a) $\Phi = 10^{14}$ cm$^{-2}$; b) $\Phi = 10^{15}$ cm$^{-2}$.

We show in Figure 9a the variation of $B_iO_i^{A(0/+)}$ TSC peak determined from consecutive measurements, performed in the same conditions, after exposing the sample for 15 minutes to ambient laboratory light. The determined concentrations are given in Figure 9b and they show that the concentration of $B_iO_i^{A(0/+)}$ is stabilized in the sixth TSC measurement, recorded at ~ 7h after the exposure to light.

The minimum and maximum concentration of $B_iO_i^A$ detected by TSC in the PAD sample irradiated with $10^{15}$ n/cm$^2$ is $1.2\times10^{13}$ cm$^{-3}$ and $2.3\times10^{13}$ cm$^{-3}$, respectively. The corresponding change in the depletion voltage is $\Delta V_{dep}$ ~ 17.6 V, large enough to alter also the evaluation of the „acceptor removal" rate, $g_B$. It is worth noting that these variations were triggered only by the a 15 minutes exposure to the ambiental laboratory light, a common situation for any experimentalist.

In addition, as the present study did not revealed an energy level that could possibly be associated to the B configuration of the $B_iO_i$ defect, the total amount of $B_iO_i$ cannot be yet estimated and so, the true introduction rate of the defect. This is because for the B configuration of the defect we only can estimate the variations in concentration and not the steady state values. In this respect the estimation of defect introduction rate through its A configuration, of $g_{BiOi}^A = 0.023$ cm$^{-1}$ in this case, is regarded to be the minimum value, as $g_{BiOi}$ should be calculated considering both defect configuration ($g_{BiOi} = g_{BiOi}^A + g_{BiOi}^B$).

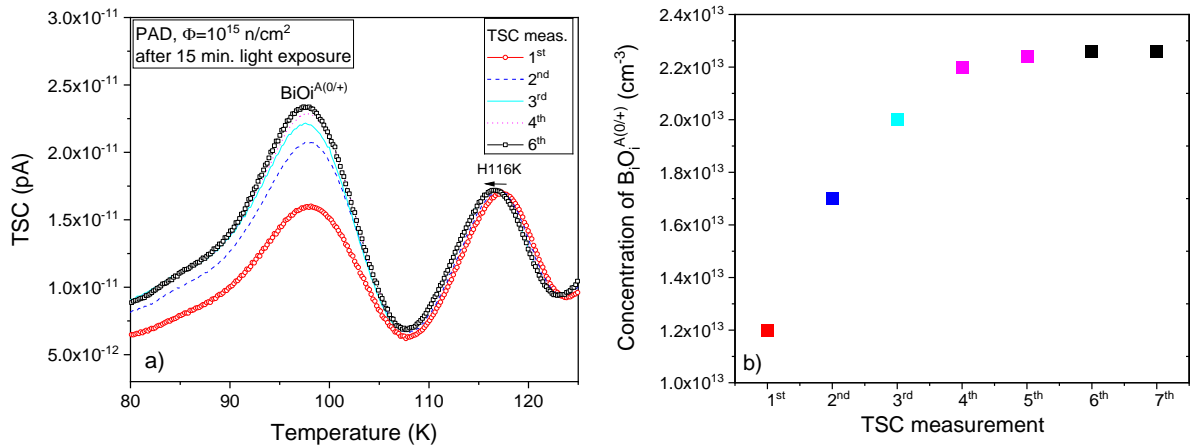

**Figure 9**. Variations in $B_iO_i^{A(0/+)}$ concentration during consecutive TSC measurements after exposure for 15 min. to ambient light: a) repetitive TSC spectra recorded for $U_R = 300$ V, after 0.3 mA forward injection at 10K; b) evaluated [$B_iO_i^A$] values.

Bistable defects have normally more than one stable state which, under certain conditions may reversibly pass over from one state to another and so change their spatial and electronic structure [46-49]. There are several examples of radiation induced metastable centers in Silicon, some with bistable character (see the [50] review), and not always the different possible electronic states of the defect could be evidenced on the same samples on which the metastability was first detected or with the same investigation techniques. Among various bistable defects, we give here as examples, the $C_iC_s$ complex [51, 52], the $V_3$ center [53-55] and the earlier thermal donors in oxygen rich silicon [56-59]. According to the literature, in order to stabilize one or the other of the two defect configurations, specific conditions have to be met (e.g. temperature, injection level, bias applied, Fermi level) and thus dedicated procedures had to be employed. The instability of the $B_iO_i$ center detected in this study, when only the common procedures have been used, could directly reveal only one of the possible defect states, the one giving rise to the peaks at 120 K and 100 K in DLTS and TSC spectra, respectively. In this A configuration, the defect has a direct and strong impact on $N_{eff}$ and the variations recorded after the inherent manipulation of the samples will result in a significant spread of the results. Regarding the not yet detected B configuration of $B_iO_i$ evidenced to exist by the present study, further studies are needed to clarify its electronic properties. We can only say that most likely it is not charged at ambient temperatures.

*C) Irradiation fluence of $10^{19}$ 1 MeV neutrons/cm² – microstructural investigations*

The HRTEM image given in Figure 10a shows the presence of some tracks normal to the film surface in the $p^+$ gain layer of the LGAD. The diffraction pattern (inset of Figure 10a) shows crystal planes corresponding to the 227 space group Si (a=5.38) under the assumption of dynamical diffraction.

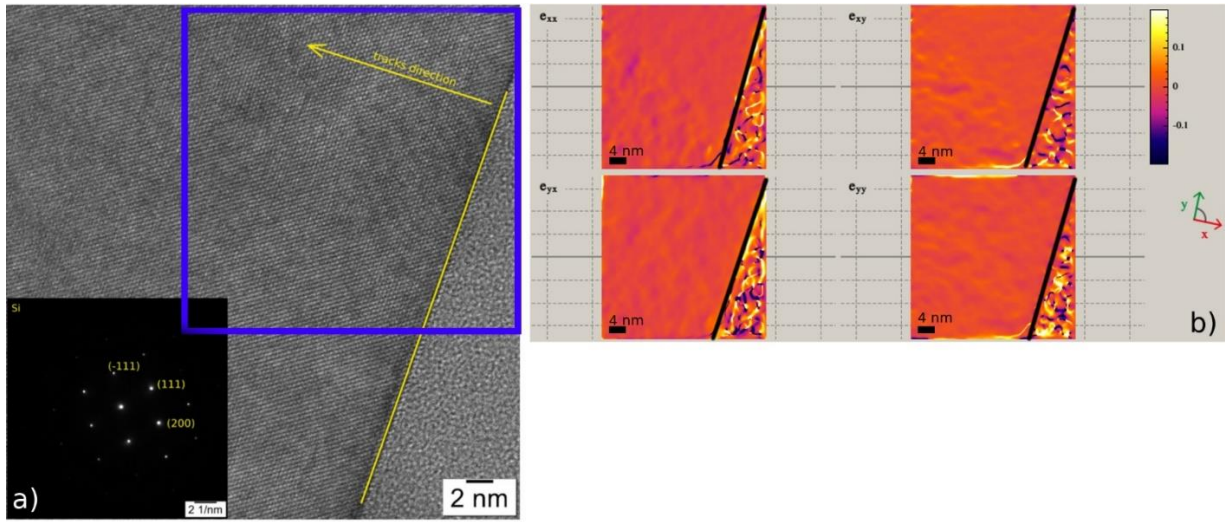

**Figure 10.** a) HRTEM image of the Si-SiO$_2$ interface. SAED showing the 227 space-group Si (inset); b) Strain maps obtained using the blue marked area in HRTEM image. The provided intensity scale represent the relative variations of the local distortions.

The presence of tracks does not provide any feature in the diffraction pattern suggesting that they are composed out of point defects. The nature of the tracks has been further investigated by using strain maps shown in Figure 10b. These have been obtained from HRTEM images as mappings of local displacements of the crystal structure relative to a free of tracks reference area. The displacements are obtained under the infinitesimal strain assumption, therefore the strain components in the images can be interpreted as displacements along or normal to the interface (exx,eyy) and shear displacements (exy,eyx). The Strain++ software has been used for the mappings [60]. The low magnitude of the strain (max 1-2% in all cases) and the uniformity of the maps, together with the HRTEM image and the SAED, prove the absence of extended defects in the crystallographic sense (dislocations, stacking faults) and suggest that the tracks are composed of defects consisting in agglomeration of vacancies and interstitials. Further, the effect of the annealing at 80 °C has been investigated from micro-structural point of view by HRTEM. For this, the LGAD-TEM specimen has been annealed in the oven at 80 °C in Ar atmosphere for 30 days. In order to exclude beam-induced artifacts, the effect of electron beam irradiation has been checked by focusing the beam for a longer than usual time (~10 sec), exposing thus the investigated material at a high electron dose. The corresponding HRTEM images are given in Figure 11. By comparing the images obtained before and after beam irradiation no significant artifacts have been observed.

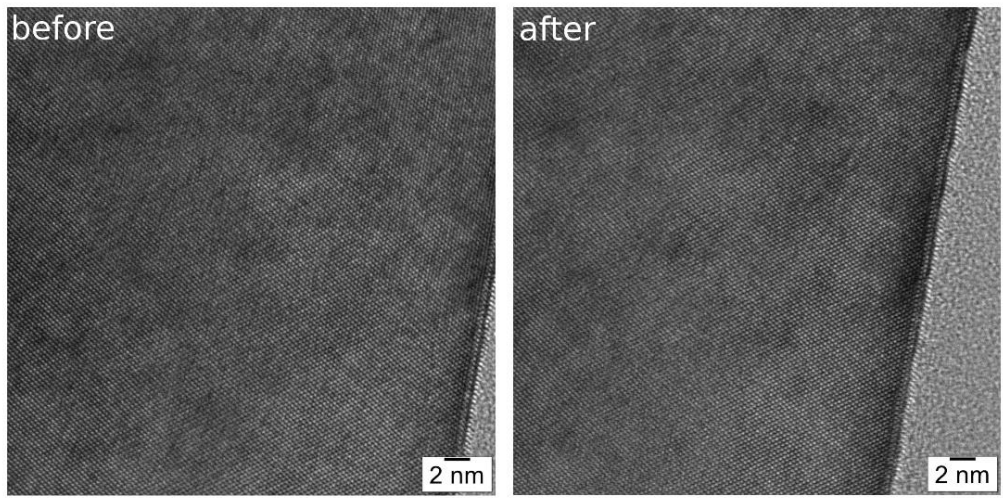

**Figure 11.** HRTEM images obtained before and after focusing e-beam on LGAD sample annealed at 80 °C for 30 days.

The HRTEM image of the sample treated at 80 °C looks similar to the one of not-annealed samples, the presence of the tracks on [001] direction, being visible in both cases. They are pointed out in Figure 12a and details can be seen in Figure 12b. The tracks show indeed agglomerations of point defects only. The inter-track distance is roughly around 1nm normal to the Si-SiO$_2$ interface. In spite of the tracks being observed both before and after thermal treatment, the defect dynamics was not revealed by HRTEM investigations.

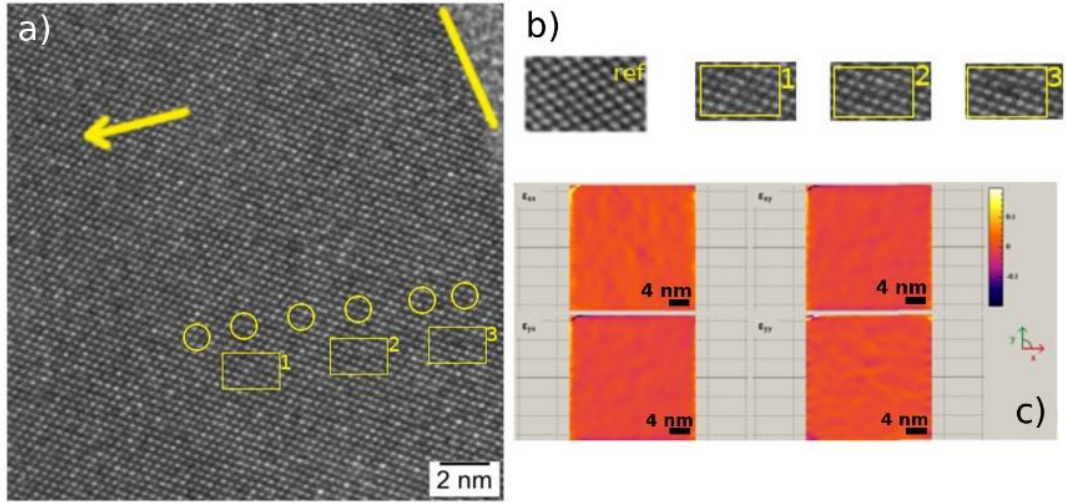

**Figure 12.** a) HRTEM image; b) details showing agglomeration of point defects as compared to the reference; c) strain maps showing the absence of dislocations, stacking faults.

The linear pattern of the defects is observed in the first tens of nm beyond the interface, on the [001] crystalline direction. Further investigations shall clarify if and how such pattern evolves deeper in the bulk. Strain maps obtained on the HRTEM image in Figure 12a confirm that extended defects such dislocations, stacking faults do not appear during the annealing at 80 °C. We associate the presence of the tracks with an interplay between the interactions of the 1 MeV neutrons with matter and the defect dynamics within the Si crystal lattice. The effects of highly energetic particles on crystalline materials are a consequence of complex physical mechanisms within the considered materials. Depending on the material, fluence, energy and nature of the irradiating particle, the effects can be observed as either point defects, extended defects and/or local amorphization of the material. The present study shows that even for very high irradiation fluences of 1MeV neutrons, the Si structure remains crystalline and the only observed imperfections are due to the agglomeration of point defects. Worth to note that although the irradiation with neutrons took place from all the directions (Triga reactor) the damage observed in the first few nanometers from the Si/SiO$_2$ interface seems to prefer the [001] crystalline direction.

## 4. Conclusions

Our study focused on the B$_i$O$_i$ defect, as determined from DLTS and TSC experiments, in connection with the acceptor removal rates in B-doped silicon diodes irradiated with 23 GeV protons and 1 MeV neutrons. Both capture cross sections, for electrons and for holes, have been directly measured for the $B_iO_i$ defect as well as its concentration in different samples via analyzing the emission from the $B_iO_i$ donor level located at about 0.25 eV from the conduction band. We followed the dependencies on doping, irradiation fluence and particle type in a try of understanding the large scattering in the results reported previously for acceptor removal rates in p type Si. The $B_iO_i$ introduction rate after proton irradiation start to be larger than after neutron irradiation for fluences $\Phi_{eq} > 10^{12}$ cm$^{-2}$. We have evidenced that even when using the same vendors and processes for all the investigated diodes and performing all the experiments in the same place, with the same set-up and procedures, the scattering of data starts to be visible for irradiation fluences above $10^{12}$ $n_{eq}$/cm$^2$ and become significant at larger fluences even when measuring the same sample.

We demonstrated that the main reason behind is the metastable behavior of the $B_iO_i$ defect which can exists in at least two configurations (labelled as A and B). The switch between the different $B_iO_i$ defect configurations evidenced in this was observed for defect concentrations exceeding $10^{12}$ cm$^{-3}$ in both, high resistivity and medium doped silicon and only through variations detected in the A configuration of the defect, characterized by a donor energy level at ~0.25 eV from the conduction band. The defect reversibly passes from one configuration to another after exposing the samples to an excess of carriers, achieved by thermal treatments at moderate temperatures or by the inherent exposure to the ambient light when manipulating the samples prior to the electrical measurements performed in dark, indicating thus a bistable character of the center. It also change its electrical activity causing not only significant long time variations in both, the effective doping concentration - $N_{eff}$ and the concentration of $B_iO_i$ defect, as determined from the emission of electrons by the donor level of the A configuration, but also an underestimation of the true introduction rate of the $B_iO_i$ defect and of the "acceptor removal" rate. Any electrical measurement performed before the $B_iO_i^{A(0/+)}$ configuration is stabilized will give a different result. Thus, we conclude that such procedural reasons are contributing significantly to the large scattering in both, the reported values concerning the "acceptor removal" as determined from C-V/I-V measurements and the $B_iO_i$ introduction rate as detected in DLTS and TSC measurements. We expect that the magnitude of variations and time constants depend on the impurity content in the samples (boron, Carbon, Oxygen), on the light intensity, on the damage status of the samples (e.g. Fermi level position in the bandgap, leakage current) and on temperature, and such dependencies shall be studied further.

Measuring the steady state values are very important for planning operation scenarios in LHC experiments (pitch dark). In addition, in order to be able to make the right correlations between generation of the defects and device performance, that are important for developing radiation hard devices based on defect engineering approach, one has to count again the steady state values and for this the best solution is to keep the irradiated samples in dark, for a given time, prior to any electrical measurement, especially in the case of highly irradiated samples. For the cases presented here, it takes about 7 h for the detectable $B_iO_i^A$ defect configuration to reach the steady state concentration. Since not always it is possible to wait such long times, further investigations are needed for parametrizing the variations in time of both, $B_iO_i$ defect concentrations and $N_{eff}$, for different situations.

An LGAD silicon sample irradiated with a very high neutron fluence has been subjected to micro-structural investigations by HRTEM by investigating the first tens of nm from the $Si/SiO_2$ interface, in the p+ gain layer of the LGAD. Point defects (vacancies and interstitials) appear to organize in tracks orthogonal to the $Si/SiO_2$ interface, on [001] crystalline direction, with an inter-track distance roughly around 1nm. The presence of the tracks is associated with an interplay between the interactions of the 1 MeV neutrons with matter and the defect dynamics within the Si crystal lattice. The potential damage of the electron beam of the TEM has been considered and the authors are confident that the observed defects are not in-situ induced artifacts. Worth noting is that even so high irradiation levels, no signs of amorphization of the Silicon crystalline structure was found.

**Acknowledgements**: This work has been performed in the framework of the CERN-RD50 collaboration (the "acceptor removal" Project) and it was financed by IFA-CERN through the Project DEPSIS and Core Program 2019-2022 (contract 21N/2019). I. Pintilie would like to thank Alexander von Humboldt Foundation for providing the DLTS system, necessary for performing the measurements.